# SIMULATION STUDIES OF THE MOLIERE RADIUS FOR EM CALORIMETER MATERIALS

*O. P. Gavrishchuk[1], V.E. Kovtun[2], T.V. Malykhina[2]*
[1]*Joint Institute for Nuclear Research (JINR), 6 Joliot-Curie St, Dubna, Moscow Region, 141980, Russia;*
[2]*V.N. Karazin Kharkiv National University, 4, Svobody sq., 61022, Kharkiv, Ukraine*
*E-mail: vladimir.e.kovtun@univer.kharkov.ua*

The Monte Carlo calculations of the Moliere radius (RM) for some homogeneous and heterogeneous media used in electromagnetic calorimetry in the energy range from 50 MeV to 10 GeV are presented in detail. The obtained results, the uncertainties in determining RM, estimations of the absorbed energy, methods for approximating the absorbed energy, and the accuracy of the results are discussed as well. Some RM are shown for calorimeter prototypes of the Spin Physics Detector experiment (SPD). A one-parameter function of the Moliere radius dependence on the absorber-scintillator thickness ratio is obtained.
PACS: 02.70.Uu, 7.05.Tp, 29.40.Vj

## 1. INTRODUCTION

Requirements for the electromagnetic ECal calorimeter emerge from the physical problems of the SPD NICA experiment [1]. A good energy resolution of 5% must be ensured as well as effective $\pi^0$-$\gamma$ separation in the energy range from 50 MeV to 10 GeV. The calorimeter module is a sampling lead-scintillator structure, which has been investigated and significantly improved for experiments KOPIO [2] and COMPASS II [3]. The polystyrene scintillator thickness is the same for all prototypes and is $D_{PS}$=1.5 mm. The main prototype of the ECaL SPD module has 4 cells and the silicon photomultiplier (SiPM) with fiber readout. The final design version will take into account the price factor and a detailed calculation by the Monte Carlo method for the improvement of the SPD calorimeter parameters [4, 5].

In this work we present the results of Monte Carlo simulation of the transverse evolution of an electromagnetic shower in the ECal SPD module. The final goal is to obtain reasonable values of the Moliere radius for the heterogeneous structure of the prototypes of the SPD calorimeter module.

## 2. $R_M$ CALCULATIONS METHODS

By definition [6], the Moliere radius $R_M$ is found from the transverse dimension of the electromagnetic (EM) shower absorbed by the medium according to the formula:

$$0.9 = \frac{E(R < R_M)}{E(R < \infty)} \quad (1)$$

The results of the numerical solution of equation (1) based on the first principles of the EM shower propagation in a medium by the Monte Carlo method [7] showed a significant difference with the estimates according to the known formulas for $R_M$ both for homogeneous media and for heterogeneous media.

We consider various methods for solving equation (1) by calorimeter modeling and subsequent processing of the obtained data.

The first standard step consists of the generation of a large number (in our case, N=10$^4$) events using the Geant4 toolkit for simulation of electromagnetic showers from electrons with primary energy of 1 GeV. The simulation of an infinite calorimeter was performed in order to avoid shower energy leakage. At the next step, 90% of the deposited energy is summed up in accordance with the chosen method.

### 2.1. RM CALCULATION FROM THE TRANSVERSE SHAPE OF THE EM SHOWER

There are many parameterizations of the transverse shape of the EM shower. The LumiCal collaboration function was chosen as an example [8]. This function was used to process experimental data from experiments specifically devoted to finding the Moliere radius. Comparisons (Fig. 1) were also made with Monte Carlo calculations.

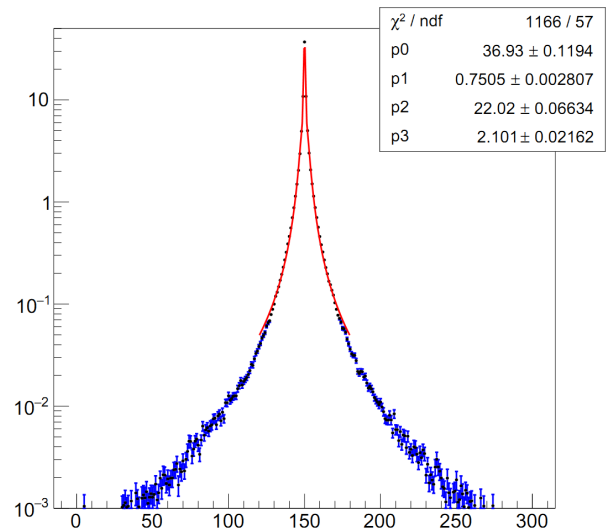

*Fig. 1. Parametrization of the shower transverse profile. The function is taken from [8].*

It was assumed that the shape of the EM shower has an axial symmetry, and the distribution function consists of a narrow Gaussian kernel and a wide hyperbola tail. The function has four parameters that are found when fitting. The result of this method depends on the region of convergence and has, as a rule, a large parameter error.

## 2.2. RM CALCULATION BY FRACTION OF DEPOSITED ENERGY

As a rule, the method for calculating the deposited energy is based on the fact that the calorimeter is divided into pads with coordinates $x_i y_j$. In our case the pad size is selected $0.15 \times 0.15$ mm$^2$. The deposited energy is equal $E_{90\%} = \Sigma E_{ij}$ provided $R < R_M$. Figure 2 shows a fragment of the geometrical calculation of the Moliere radius for some homogeneous materials that are used in calorimetry.

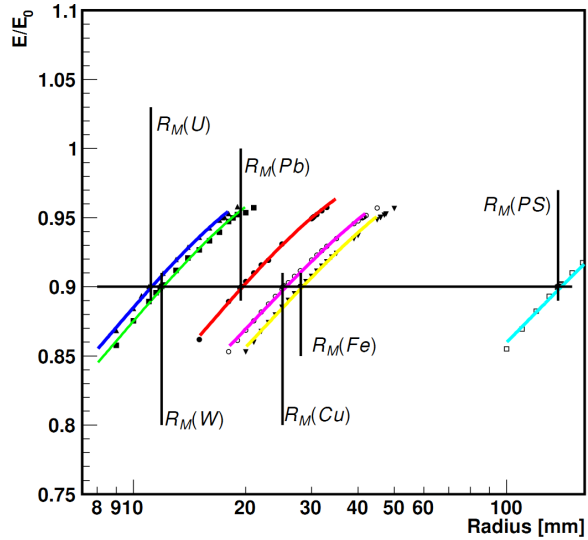

Fig. 2. Geometric method of $R_M$ calculation for some homogeneous media.

The Moliere radius is determined by the intersection of the E=E(R) function and the straight line $E = 0.9 \cdot E_0$, using this method. Table 1 presents the numerical values of the Moliere radius for some homogeneous materials. The presented results were obtained by the method described above.

**Table 1.**
*Moliere radii for some homogeneous materials*

| Material | Pb | W | U | Cu | Fe | PS |
|---|---|---|---|---|---|---|
| $R_M$, mm | 19.4 | 11.9 | 11.14 | 25.1 | 28.1 | 137.5 |

The systematic error of the method is obviously determined by the pad size and in our case is equal $\Delta R_m(\text{syst.}) \sim \pm 0.2$ mm.

## 2.3. RM CALCULATION FROM ENERGY SPECTRUM

We propose a method for calculating $R_M$ directly from the 90% peak of the total absorption of an EM shower in the cylindrical volume of the calorimeter. An example of an energy spectrum is shown in Fig. 3.

In this case, the shape of the peak will differ from the Gaussian due to the energy leakage of the shower into the outer cylinder.

The total absorption peak has the form of a δ-function under the condition $R < \infty$ since $E_{mean} = E_0$. Thus, the problem is reduced to the correct determination of the peak maximum when 90% of the shower energy is absorbed. It should be noted that this effect of volume energy leakage also exists for the previous case of calculations. This effect is usually ignored when calculating the Moliere radius. The question is what value of the deposited energy should be chosen. Whether the most probable value of the deposited energy or the average deposited energy value should be chosen.

In our case ($E_0$=1 GeV), the difference between the most probable energy distribution of events and the average value is ~4 MeV (0.5% of $E_0$). The difference is not very large for practical use, so we use the most probable value.

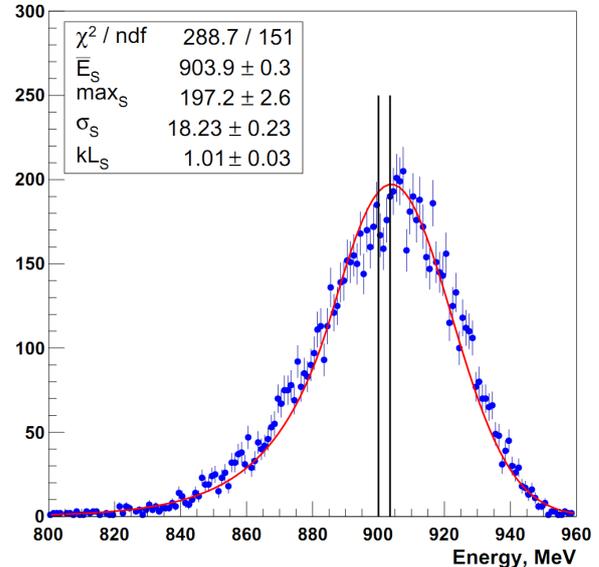

Fig. 3. Energy spectrum of 90% of the deposited energy in a lead cylinder. Vertical lines are the average and the most probable value of deposited energies.

The peak was approximated by the Das function [9], from which the distribution parameters were obtained. The most probable peak value $E_S$ and the standard deviation σ are shown in Fig. 3.

Thus, we consider the most probable value of the resulting distribution of the deposited energy in the cylinder, and not the average value. Such energy spectra are close to the Gaussian distribution. Therefore, we have $\chi^2/\text{ndf} \sim 1$ for the used approximation as a rule [9].

The difference in the results of the two previous methods is negligible, so the subsequent results were obtained by the pads method.

## 3. $R_M$ FOR IDEAL SPD ECAL

The obtained values of Moliere radii for ideal prototypes of calorimeters with Pb and W absorbers are presented in Tables 2 and 3.

**Table 2.**
*Moliere radii for ideal prototypes of calorimeters with Pb absorbers*

| $D_{Pb}$[mm]+$D_{PS}$(1.5 mm) | 0.3 | 0.4 | 0.5 |
|---|---|---|---|
| $R_M$ [mm] | 74.0 | 65.0 | 58.5 |

**Table 3.**
*Moliere radii for ideal prototypes of calorimeters with W absorbers*

| $D_W$[mm]+$D_{PS}$(1.5 mm) | 0.3 | 0.4 | 0.5 |
|---|---|---|---|
| $R_M$ [mm] | 53.2 | 45.3 | 40.0 |

$D_{PS}$, $D_{Pb}$, $D_W$ – thickness of the active and passive parts of the calorimeter.

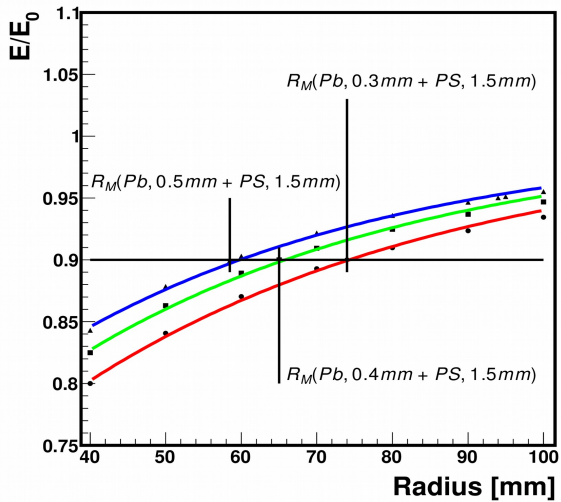

*Fig.4. Geometric method of $R_M$ calculation for ECal prototypes with lead absorber*

We present the result of $R_M$ calculations for the heterogeneous structure of the SPD calorimeter prototype (Fig. 4) for comparison with the homogeneous structure (Fig. 2).

## 4. RM DEPENDENCE ON THE ABSORBER THICKNESS

The Monte Carlo method was used to investigate the Moliere radius in more detail, depending on any ratio of the thicknesses of the passive and active substances of the calorimeter. A universal approximation function is found:

$$R_M(x) = R_M^{PS} \cdot \frac{1-x}{1+a \cdot x} + R_M^{abs} \cdot \frac{x + a \cdot x}{1+a \cdot x}, \quad (2)$$

where the variable x varies within [0,1]:

$$x = \frac{D_{abs}}{D_{abs} + D_{PS}}, \quad (3)$$

$$R_M(0) = R_M^{PS}, \quad R_M(1) = R_M^{abs} \quad (4)$$

$D_{PS}$, $D_{abs}$ – the thicknesses of the active and passive parts of the calorimeter. $R_M^{PS}$, $R_M^{abs}$ - Moliere radii of the plastic scintillator and absorber, $R_M(x)$ - the Moliere radius of the sampling calorimeter as a function of the ratio of the thicknesses x. The 'a' parameter is the only dimensionless fitting parameter in the formula (2).

Fig. 5 shows functions (2)-(4) for a sampling calorimeter with lead and tungsten absorbers and the 'a' parameter values obtained by fitting.

## 5. THE RESULTS TESTING

In order to obtain the test results and identify systematic errors, the parameters of computer models were varied. Figure 6 shows that in a wide range of RangeCut values (from 1 mkm to 1 mm) the $R_M$ calculation results do not differ significantly. The calculations were also performed with a statistical accuracy of 1% for incident electrons or photons in energy range from 100 MeV to 10 GeV.

During a series of calculations, the most suitable model of physical processes emstandard_opt4 was chosen instead of QGSP_BERT model in the PhysicsList class.

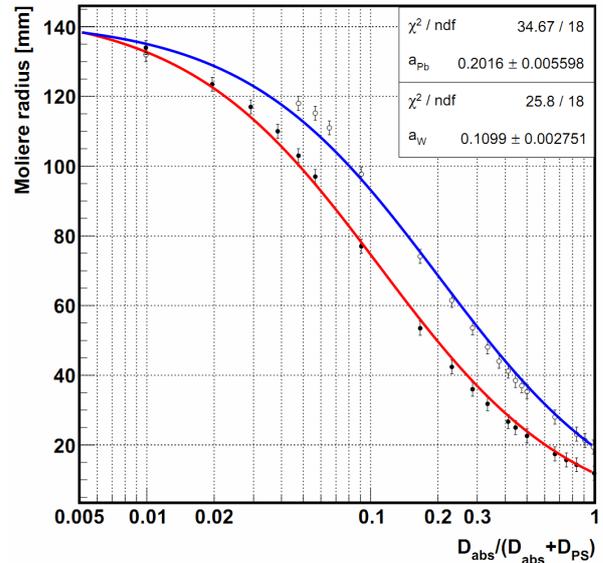

*Fig.5. Universal function from (2)-(4) and $R_M$ dependence on variable thickness x and fitting parameter 'a'.*

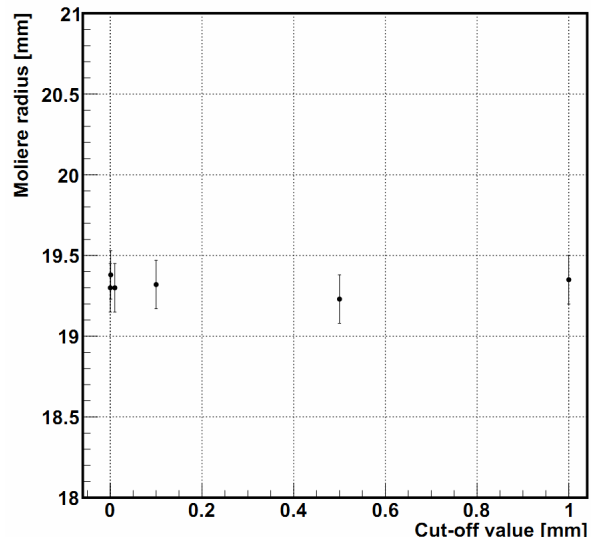

*Fig.6. $R_M$ dependence on the cut-off value.*

This allows to focus on purely electromagnetic processes without taking into account nuclear reactions, which appear in a small quantity when using QGSP_BERT model. It is also shown (Fig. 7) that the Moliere radius does not depend on the energy of the incident particle.

The increase of $R_M$ in function $R_M=R_M(E_0)$ at the beginning of the curve can be explained by the actual absence of multiple processes of an electromagnetic shower in the region of very low energies. It is also difficult to find the correct definition of $R_M$ in this area.

Errors in $R_M$ calculations are determined by the statistical error associated with the number of events and systematic errors due to data processing methods.

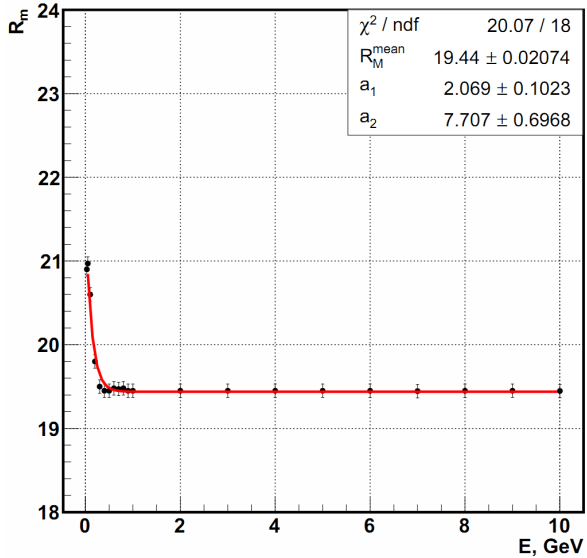

*Fig.7. RM dependence on the energy of primary particles.*

To determine the latter, series of calculations were carried out. We estimate the total errors in calculating $R_M$ as $\Delta R_M \pm 0.5$ mm for the data from Tables 1, 2, 3.

## 6. CONCLUSION

A simulation study of the Moliere radius for an ideal sampling calorimeter is presented. Various practical methods are shown for solving the equation for the Moliere radius, which follows from the Moliere radius definition. The data were obtained by the Monte Carlo method. A convenient approximation of the curve of the Moliere radius dependence on any possible thicknesses of the active and passive parts of the sampling calorimeter is found. Thus, the formula is suitable for use in both homogeneous and heterogeneous environments. An estimate of the calculation accuracy for the obtained results is made, which can be used in the development of a sampling calorimeter of the SPD NICA setup. The obtained results are practically independent of some Geant4 parameters, such as cut, the Physics Lists model, and the energy range of electrons or gamma quanta. The methodology described in this paper for Moliere radius calculation can be easily adapted to any sampling calorimeter.

**SIMULATION STUDIES OF THE MOLIERE RADIUS FOR EM CALORIMETER MATERIALS**

*O. P. Gavrishchuk, V.E. Kovtun, T.V. Malykhina*

The Monte Carlo calculations of the Moliere radius ($R_M$) for some homogeneous and heterogeneous media used in electromagnetic calorimetry in the energy range from 50 MeV to 10 GeV are presented in detail. The obtained results, the uncertainties in determining $R_M$, estimations of the absorbed energy, methods for approximating the absorbed energy, and the accuracy of the results are discussed as well. Some $R_M$ are shown for calorimeter prototypes of the Spin Physics Detector experiment (SPD). A one-parameter function of the Moliere radius dependence on the absorber-scintillator thickness ratio is obtained.